# Highly polarized single photon emitter from intrinsic localized excitons in a WSe$_2$/CrSBr heterostructure.

*Varghese Alapatt[1], Francisco Marques-Moros[1], Carla Boix-Constant[1], Samuel Manas-Valero[1,2], Kirill Bolotin[3], Josep Canet-Ferrer[1]\*, and Eugenio Coronado[1]\**

[1]Instituto de Ciencia Molecular (ICMol), Universitat de València, Paterna, Spain.

[2]Kavli Institute of Nanoscience, Delft University of Technology (TU Delft), Delft, Netherlands.

[3]Department of Physics, Freie Universität Berlin, Germany.

## ABSTRACT

Single photons emitters (SPEs) are key components in quantum information applications and are commonly generated in 2D materials by inhomogeneous strain engineering. Here, we report an alternative approach that involves a 2D semiconductor/2D magnet heterostructure. The optical study of the WSe$_2$/CrSBr heterostructures reveals several new emission lines at lower energies compared to characteristic WSe$_2$ emissions, that are assigned to localized excitons. Further investigation demonstrates that one of these emergent lines is an SPE with a strong valley polarization response and large energy shift with the field-induced metamagnetic transition in CrSBr, linking it to the magnetic proximity effect of the adjacent CrSBr layer. In contrast to previous reports on WSe$_2$ that only allow tuning of the SPEs by out-of-plane magnetic field, our emitter is sensitive to both in- and out-of-plane fields. Our findings demonstrate the potential of this approach for improved control and polarization of SPEs in 2D materials.

INTRODUCTION

Single photon emission of localized excitons is a fundamental constituent in quantum information technologies[1,2] and quantum photonics.[3,4] Atomic defects[5] and localized strain in monolayer WSe$_2$ have been extensively reported to facilitate the localization of excitons that exhibit quantum emitter properties.[6-9] In addition to being easy to integrate with other nanostructures,[10-12] monolayers of WSe$_2$ also afford an additional valley degree of freedom due to the presence of inequivalent K valleys in their electronic structure.[13,14] The flexibility of this 2D material facilitates the strain engineering of local potential wells. Such an approach is the most successful to generate single photon emitters (SPEs) in 2D systems. However, a crucial requirement for the SPEs to be integrated with optical cavities is to have pure and controllable polarization of the emission.[1,15,16] While rapid progress is being made towards achieving this,[17] the strain-induced SPEs' polarization is mostly dependent on the nature and orientation of the strain tensor,[18,19] being the valley response often altered.

A promising alternative could be the integration of WSe$_2$ and similar semiconducting transition metal dichalcogenides (TMDs) with 2D magnetic materials[20-24] since they can modify or enhance their optical response to electric fields,[25] magnetic fields,[24,26] or mechanical strain.[27]

Popular among them are the ones having an intrinsic out-of-plane magnetic anisotropy, like CrI$_3$,[28-30] as they can induce spontaneous valley splitting and enhanced valley polarization.[28,31,32] However, this approach has not yet been exploited to generate SPEs. In this work, we explore the use of the in-plane 2D magnet CrSBr to reach this goal. This 2D A-type antiferromagnet is gaining interest[24,33] due to its high in-plane anisotropy[34,35] and the presence of the same spin-selective magnetic centers (Cr$^{3+}$) as in CrI$_3$[26,28] and CrBr$_3$.[37,38] Here we will show that this WSe$_2$/CrSBr semiconducting/magnetic heterostructure can be exploited to generate highly polarized single photons that are coupled to the magnetic states of CrSBr. Since CrSBr lacks any spontaneous out-of-plane magnetic moment[38-41] this approach can also be promising for developing applications that require fast perpendicular magnetic switching/sweeping.

RESULTS

To construct the heterostructure, a monolayer of WSe$_2$ is placed on top of a few-layer (6 layers) CrSBr and held together by van der Waals interaction (Figure 1A). Figure 1B shows the corresponding colour map of the sample, where the colour bar represents the integrated intensity of the micro-photoluminescence ($\mu$-PL) signal. In the top and bottom panel of Figure 1B the contributions from CrSBr and WSe$_2$ are shown respectively by filtering in their corresponding band gap energies (1.34 eV and 1.63 eV respectively). It becomes clear from the $\mu$-PL maps, that the PL contribution from WSe$_2$ is significantly brighter in the heterostructure region, while that from CrSBr is not significantly altered.

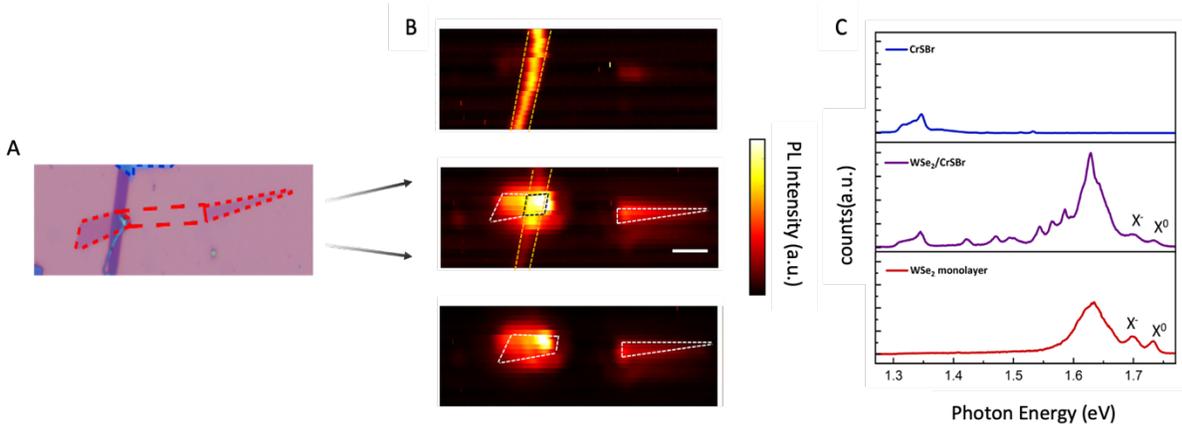

**Figure 1**. Sample configuration of WSe$_2$/CrSBr Heterostructure. (a) Optical microscope image of the WSe$_2$/CrSBr heterostructure over a silica on silicon substrate (scale is 5 $\mu$m). The rectangular purple flake is a few-layer CrSBr flake(~6 layers, determined by optical contrast), and the flake outlined by red dashed line is a monolayer of WSe$_2$ placed on top of it. Note that the right WSe$_2$ is separated and can be used as reference. (b) Colour map of the PL integrated intensity of the sample at low temperature (~10K). The top and bottom panels represent the corresponding maps filtering just the contribution from CrSBr and WSe$_2$, respectively and the middle panel represents the map without any filtering. The reference layers (CrSBr flake and WSe$_2$ monolayer) and the vdW heterostructure are marked by yellow, white and black dashed lines, respectively. The scale is 5 $\mu$m and is the same for all the maps (c) Representative PL spectra from CrSBr reference (top), WSe$_2$/CrSBr heterostructure (middle), and WSe$_2$ reference (bottom), all at 60K.

Figure 1C presents a representative µ-PL spectrum from CrSBr, WSe$_2$ (top and bottom panels respectively) and the WSe$_2$/CrSBr heterostructure (middle panel). The spectra at 60 K are chosen so that the characteristic neutral exciton (X$^0$~1.73 eV) and trion (X$^-$~1.7 eV) emission peaks[42,43] start to emerge in µ-PL spectra of the heterostructure (see Figure SI 4). The relatively broad emission band of WSe$_2$ at lower energy is generally attributed to localized excitons (LEs).[43,44] There is a noticeable decrease in the intensity of neutral exciton and trion emissions of WSe$_2$ in the heterostructure accompanied by a significant rise in the intensity of the LE emissions. At the interface with CrSBr, LEs in WSe$_2$ seem to dominate over other species. However, a more striking feature in the µ-PL of the heterostructure is the emergence of several new peaks at lower energies compared to that of the WSe$_2$ main band. These peaks are also attributed to LEs formed in the WSe$_2$ monolayer as they are energetically closer to the WSe$_2$ main band and display valley Zeeman splitting with an applied perpendicular magnetic field (Figure SI 3).

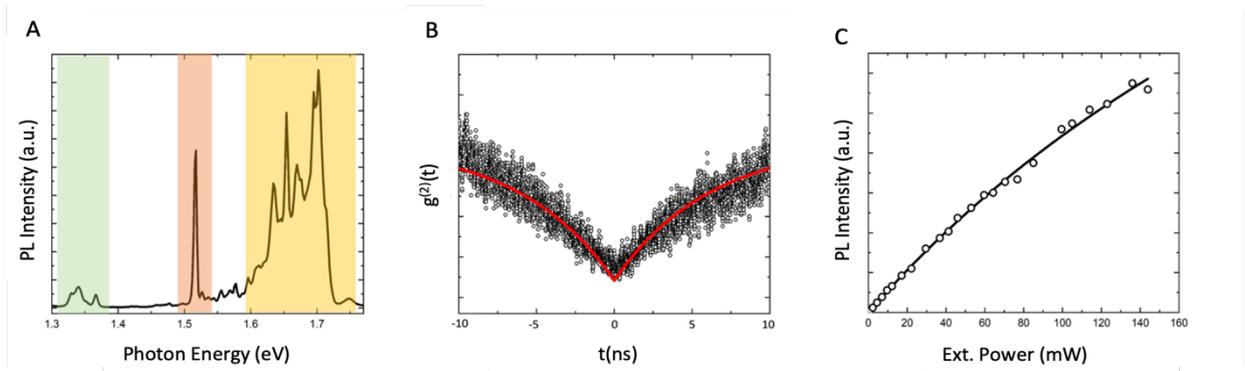

**Figure 2.** Optical properties of the emission peak at 1.51 eV (a) µ-PL spectra at ~6K from the region of the heterostructure. The shaded light green, orange, and yellow region represents the contribution to emission from few-layer CrSBr, the SPE and WSe$_2$ monolayer, respectively. (b) Second-order photon correlation measurement of the SPE measured at 6K. The scatter points represent the experimental data, and the red line is the fitted curve to the second-order photon correlation function ( $g^{(2)}(t) = 1 - (1-a)e^{-\frac{|t|}{\tau}}$ ). (c) µ-PL intensity as a function of the incident laser power. The scatter points are experimental data, and the black line is the data fitted to function $I = I_{sat} \frac{P}{P+P_{sat}}$.

The spectrum corresponding to the region of the heterostructure exhibiting the strongest LE at 6 K is shown in Figure 2A (for location see Figure SI 1). One notices that, apart from the contributions from the CrSBr and WSe$_2$ (marked by light green and yellow shaded areas, respectively), an isolated sharp peak clearly emerges at ca. 1.51eV (shaded light orange). This peak exhibits photobleaching at low temperature, which is common for highly localized excitons in 2D materials.[5,45,46] Hence, a second-order photon-correlation measurement is carried out for this emission to investigate the single photon emission behavior (figure 2B). The experimental data are fitted using the second-order photon-correlation function given by $g^{(2)}(t) = 1 - (1-a)e^{-\frac{|t|}{\tau}}$, where $t$ is the delay time between simultaneous counts, $\tau$ includes pumping and recombination rates, and $a$ gives the value of the function at t=0. The value of $g^{(2)}(0)$ extracted from the fit is ca. $0.12 \pm 0.04$, which is less than 0.5, thus confirming a pronounced photon antibunching behaviour,[7-9] characteristic of a single photon emitter. Single quantum emitters in WSe$_2$ monolayers originating from strain and defects almost always result in the SPEs occurring at an energy range between 1.6 and 1.75 eV.[18,19,47] Interestingly, in our case the SPE occurs at much lower energy and is very well separated from the characteristic WSe$_2$ emissions. This indicates the key role played by the adjacent CrSBr magnetic layer in inducing the localization of excitons in the WSe$_2$ layer. The state amplitude exhibits typical sublinear power dependence that substantiates the localized character of the excitons responsible for this emission (Figure 2C).[6,45,46] Fitting with $I \approx P^a$ we get the value of $a \approx 0.84$, also a signature consistent with SPE. It is worth to mention that the usual SPEs in WSe$_2$ gives smaller power exponents, in the range of 0.5-0.6.[48] We have focused on this SPE as it is the brightest and the most isolated one. Still, we cannot discard the presence of other SPEs among the emergent localized excitons that appear all over the heterostructure region. In fact, we found peaks at higher energies (close to that one reported for strained WSe$_2$) with the same valley dependence discussed later but not isolated enough for measuring an antibunching.

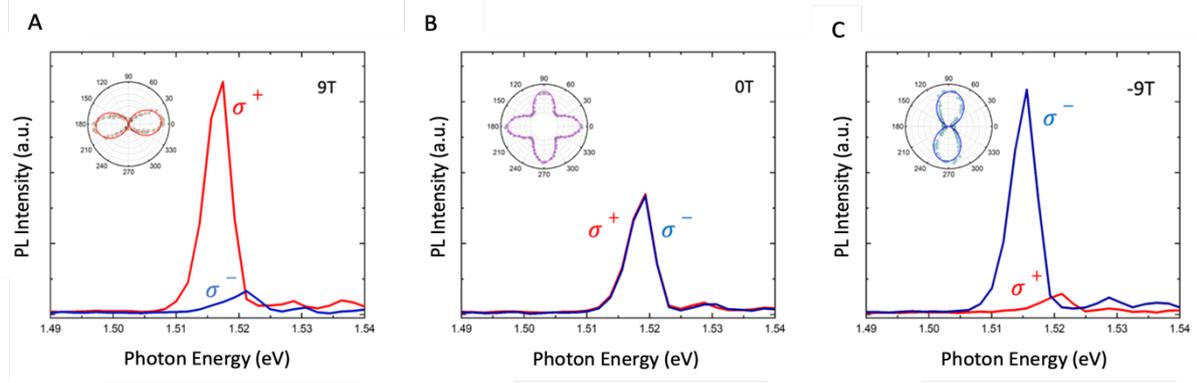

**Figure 3.** Polarization resolved $\mu$-PL of the SPE. (a), (b), and (c) Circular polarization resolved spectra of the SPE for right ($\sigma^+$, in red) and left ($\sigma^-$, in blue) circular polarizations at different external perpendicular magnetic fields of +9, 0 and -9 T respectively. The polarization dependence of the SPE $\mu$-PL emission at 1.51eV is shown as an insert for each field.

In order to demonstrate the polarization selectivity of this SPE, we carry out a polarization resolved $\mu$-PL at varying magnetic field directions. In a first step, we apply perpendicular magnetic fields (Figure 3A-C). The excitation laser source is linearly polarized while the right ($\sigma^+$, from K valley) and left ($\sigma^-$, from K' valley) circular polarizations are selected at the collection channel. At zero field we observe that the contributions from both valleys are nearly similar as shown in Figure 3B. In turn, we observe a strong preference for $\sigma^+$ and $\sigma^-$ for the largest positive (9T) and negative (-9T) fields, respectively. This result is very compelling for two main reasons: First, because having highly polarized SPEs is often challenging as the degree of polarization of strain induced SPEs in 2D TMDs are highly dependent on the nature and orientation of the strain wells created.[18,19] Hence, even though the strain induced potential wells offer precision- positioning of SPEs, an understanding of the mechanism of exciton localization and valley selection in our heterostructure contributes towards an improved control over the emitter polarization. Second, because this tuning of the valley polarization with the magnetic field is typical for charged species, such as trion and biexciton in WSe$_2$ monolayers.[42,49]

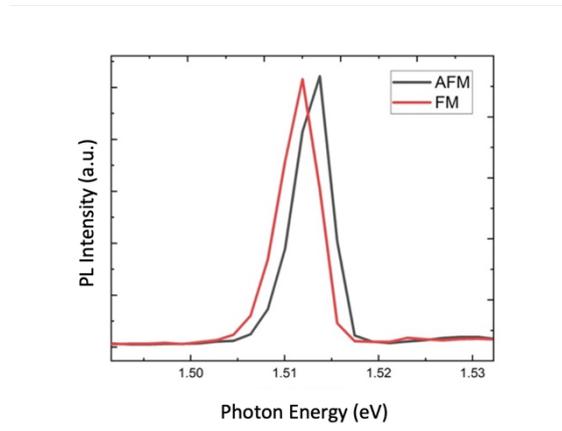

**Figure 4.** Response of PL spectra of SPE to the magnetic transition of CrSBr induced by an in-plane magnetic field. The field is applied along the easy axis of CrSBr, and the red and black spectra indicates the SPE peak when the adjacent CrSBr layer is in ferromagnetic (FM) and antiferromagnetic (AFM) states, respectively.

In an earlier work on defect related quantum emitters in WSe$_2$ a similar polarization response was observed.[5] Indeed, most parameters of the observed SPE, including the power exponent and g-factor value (*ca.* 9.8), are close to what was previously observed for defect-induced WSe$_2$ emitters. However, as mentioned before, the energy of this SPE is very well separated from the energy range in which the SPEs in WSe$_2$ usually appear. This sharp shift in energy is consistent with the screening[50] of the SPEs by the free electrons of adjacent CrSBr, thus reducing the binding energy of the SPE. In fact, CrSBr has a high dielectric constant[51] hosting a gas of free electrons.

In a second step, we perform in-plane magnetic field measurements. Figure 4 shows that SPE shifts in energy when the field-induced spin flip from AFM to FM state occurs in CrSBr. In fact, the SPE sharply shifts by about 2 meV toward lower energy when the metamagnetic transition occurs (at 0.4 T in the easy axis, and at 0.65 T in the intermediate axis). Interestingly, the magnetic fields required for reaching this shift in WSe$_2$ monolayers due to Zeeman splitting are much higher (ca. 9 T). In addition, we also observe that the intensity of WSe$_2$ peaks depends on the magnetic state of CrSBr (Figure SI 6). These observations are once again consistent with the screening effect mentioned earlier, as changes in the band structure upon the spin flip from AF to FM-like states results in a redistribution of charge across the heterostructure interface. Thus, in the FM state the carrier mobility increases as these carriers are able move among adjacent layers without reversing their magnetization. This in turn increases the number of free electrons available at the WSe$_2$/CrSBr interface thereby enhancing the screening, leading to a further decrease in the binding energy of the exciton.[52] Finally, the temperature dependent measurements show that the SPE continues to emit until moderately high temperatures (100K) before thermalizing (Figure SI 5) and persists over multiple heating/cooling cycles. This, once again, indicates a very strong localization.

CONCLUSION

We have demonstrated here how a van der Waals heterostructure interfacing a monolayer of WSe$_2$ and a few-layer 2D magnet with in-plane anisotropy, CrSBr, results in appealing emergent magneto-optical properties. In particular, we have shown, first, that several new localized excitonic states in WSe$_2$ have emerged in presence of adjacent CrSBr, with one of them behaving as a highly polarized single photon emitter (SPE). The origin of this SPE has been ascribed to intrinsic defects of the WSe$_2$. Second, we have demonstrated that in this heterostructure this quantum emitter is coupled to the CrSBr magnet, thus opening a new avenue for observing and manipulating interesting quantum phenomena. In fact, the magnetically-induced spin flip in this metamagnet can induce a 2 meV shift in the emission energy of the SPE. This contactless magnetic approach is advantageous with respect to the typical approach that needs to apply strain or electrical gaiting. Third, and in contrast to previous reports on WSe$_2$ that only allow tuning of the SPEs by out-of-plane magnetic field, our emitters are sensitive to both in- and out-of-plane fields. This approach opens then the possibility to manipulate the SPEs in any direction, including the in-plane ones. In addition, the absence of out-of-plane magnetic moment in CrSBr can also be advantageous in systems than require fast perpendicular magnetic switching, in contrast to its predecessors like $CrI_3$ with strong perpendicular moments.

More functionalities can also be induced in similar vdW heterostructures taking advantage of the in-plane magnetic anisotropy of CrSBr, as for example, the coupling of PL intensity to the magnetism of CrSBr. Our results uncover the potential of 2D magnet CrSBr which, in combination with local strain engineering, could generate highly polarized and magnetically coupled SPEs that can operate under a fast magnetic switching due to the absence of an out-of-plane magnetic moment in CrSBr. Therefore, this heterostructure opens a promising avenue to explore quantum information applications and developing functional quantum photonic devices.

METHODS

The thickness of the studied flakes was identified with an optical microscope by a calibrated contrast comparison method. (14) Micro-photoluminescence measurements were carried out on a µ-PL confocal system coupled to an attoDRY1000 cryostat and provided with a set of superconducting magnets. Inside the cryostat, the samples were placed on top of a piezoelectric three-axial stage for the accurate positioning of single flakes. A laser diode at a 520 nm wavelength with a power control driver was used as the excitation source. The diode emission was coupled to a monomode optical fiber to obtain a diffraction-limited spot with a power ranging from 0 to 100 µW. Then, backscattering was collected by means of a monomode optical fiber filtering the laser to obtain the µPL signal, which was analyzed with cooled silicon back-thinned CCD attached to a spectrometer.

ASSOCIATED CONTENT

**Supporting Information**: Additional experimental details and measurements (DOC).

AUTHOR INFORMATION

**Corresponding Authors**

**\*Eugenio Coronado** - *Instituto de Ciencia Molecular (ICMol), Universitat de València, 46980, Paterna, Spain*;

ORCID iD: https://orcid.org/0000-0002-1848-8791;

Email: eugenio.coronado@uv.es

**\*Josep Canet-Ferrer** - *Instituto de Ciencia Molecular (ICMol), Universitat de València, 46980, Paterna, Spain*;

ORCID iD: https://orcid.org/0000-0002-7221-1873;

 Email: jose.canet-ferrer@uv.es

**Authors**

**Varghese Alapatt** - *Instituto de Ciencia Molecular (ICMol), Universitat de València, 46980, Paterna, Spain;* https://orcid.org/0009-0009-9605-3273

**Francisco Marques-Moros** - *Instituto de Ciencia Molecular (ICMol), Universitat de València, 46980, Paterna, Spain;* https://orcid.org/0000-0001-8199-5326

**Carla Boix-Constant** - *Instituto de Ciencia Molecular (ICMol), Universitat de València, 46980, Paterna, Spain*; https://orcid.org/0000-0003-3213-5906

**Samuel Mañas-Valero** - *Instituto de Ciencia Molecular (ICMol), Universitat de València, 46980, Paterna, Spain*; https://orcid.org/0000-0001-6319-9238

**Kirill I. Bolotin** - *Department of Physics, Freie Universität Berlin, Arnimallee 14, Berlin 14195, Germany*; Email: kirill.bolotin@fu-berlin.de

**Author Contributions**

The manuscript was written through contributions of all authors. All authors have given approval to the final version of the manuscript.

# Supporting information for

# Highly polarized single photons from intrinsic localized excitons in a WSe$_2$/CrSBr heterostructure.

*Varghese Alapatt[1], Francisco Marques-Moros[1], Carla Boix-Constant[1], Samuel Manas-Valero[1,2], Kirill Bolotin[3], Josep Canet-Ferrer[1] \*, and Eugenio Coronado[1]\**

[1]Instituto de Ciencia Molecular (ICMol), Universitat de València, Paterna, Spain.

[2]Kavli Institute of Nanoscience, Delft University of Technology (TU Delft), Delft, Netherlands.

[3]Department of Physics, Freie Universität Berlin, Germany.

Corresponding Authors: eugenio.coronado@uv.es, jose.canet-ferrer@uv.es

**I. PL spectra of WSe$_2$/CrSBr heterostructure at 4K and spatial mappings for selected emissions.**

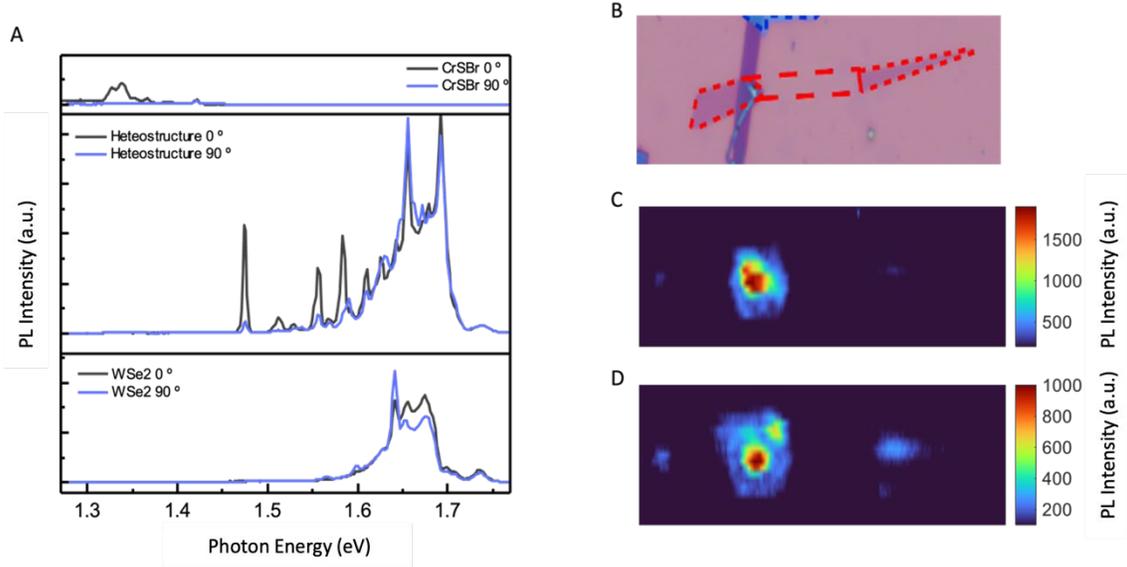

**Figure SI 1**. Basic PL characterization of the WSe$_2$/CrSBr heterostructure at 4K. (a) Representative PL spectra from CrSBr reference (top), WSe$_2$/CrSBr heterostructure (middle), and WSe$_2$ reference (bottom), all at 4K. Note that the spectra from the heterostructure is from the same point as in Fig 1(c) of the main article. Black and blue spectra represent the linear polarization resolved PL parallel to and perpendicular to the easy axis of CrSBr respectively. (b) Optical microscopic image of the heterostructure as described in the article. (c) The PL map filtering just the contribution from the isolated localised exciton emission at 1.47 eV. (d) The same plot as (c) but for the SPE signal at 1.51eV.

We observe that the characteristic neutral exciton and trion emission of WSe$_2$ are not very prominent at very low temperature. Also, we see from the energy selected PL mappings that the localized states are indeed localized mostly within the heterostructure whereas the SPE displays the very high degree of localization.

## II. Excitation power dependent PL of WSe$_2$/CrSBr heterostructure.

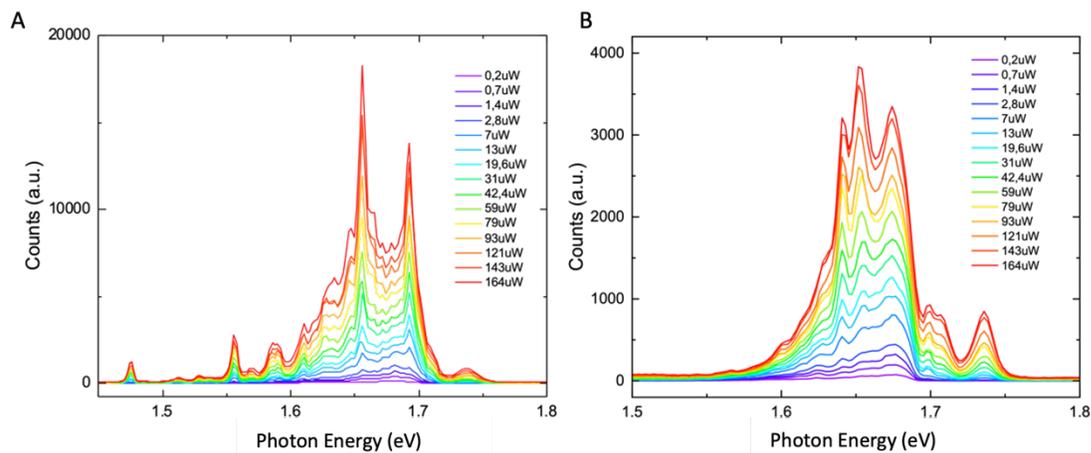

**Figure SI 2**. Excitation power dependent PL of the WSe$_2$/CrSBr heterostructure at 4K. (a) and (b) show the PL signals from the heterostructure and reference WSe$_2$ flakes at varying excitation powers. Note that the point measured here on the heterostructure is different from the one shown in FIG. S1.

## III. Zeeman splitting in perpendicular magnetic field

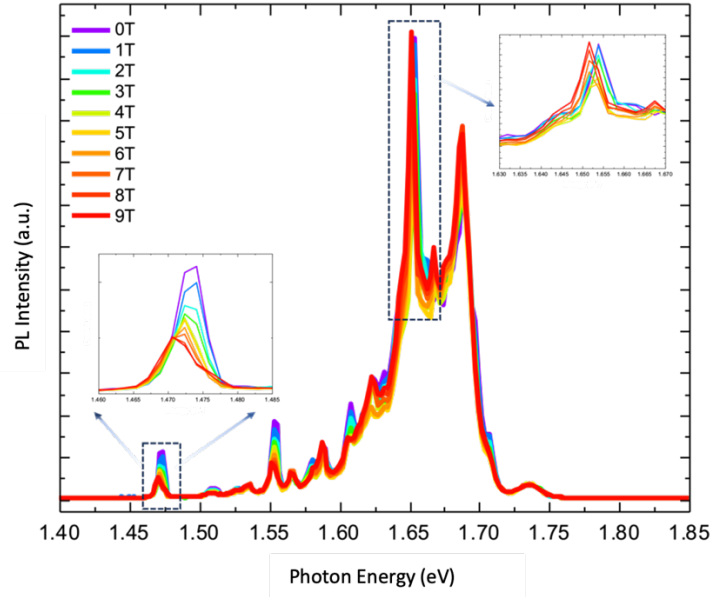

**Figure SI 3**. Out of plane magnetic dependent PL plot from 0 to 9T. The energy difference between the red and the purple plot is 2meV which is the reported Zeeman split for 9T.[1] Plots for the isolated peak at 1.47 and WSe$_2$ main band emission at 1.65eV are highlighted as insets.

The localized excitons also exhibit similar Zeeman splitting indicating that these excitons are generated from the transitions in the WSe$_2$ band which possibly are then confined by defects in CrSBr layer. Although these results allow us to reveal the origin of the LEs, we cannot conclude where they live (in WSe$_2$, in CrSBr or shared between the two materials). Further measurements, like Fourier back focal plane imaging, needs to be done to answer this question.

## IV. Temperature dependent PL of WSe$_2$/CrSBr heterostructure.

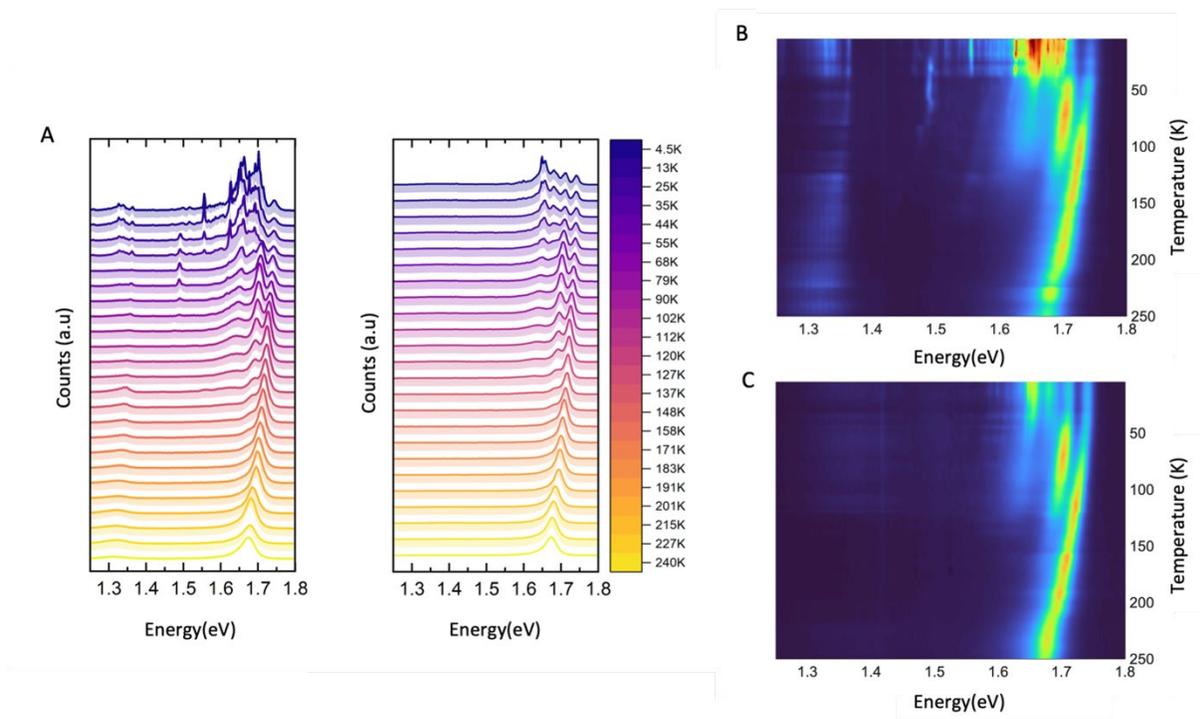

**Figure SI 4.** Temperature dependent Pl measurements. (a) Temperature-dependent PL spectra for heterostructure (left) and the reference WSe$_2$ monolayer (right) from temperatures from 4.5 to 250K (shown as colour bar). (B) and (C) Colour map composed by the spectra from (a). Colour bar represents the integrated intensities of the PL emissions.

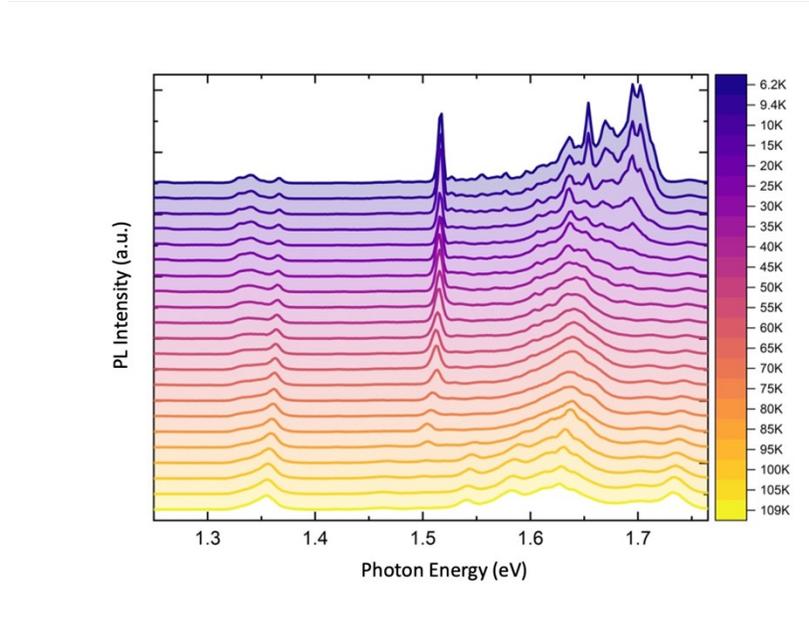

**Figure SI 5**. Temperature-dependent PL spectra from temperatures from 4.5 to 250K for heterostructure where the SPE occurs. Colour bar represents the integrated intensities of the PL emissions.

We observe that the emergent spectral features in the heterostructure gradually disappear starting from around 40K and the characteristic $WSe_2$ excitonic peaks recover their intensities and then follow the similar temperature dependence as in the reference $WSe_2$[2,3] beyond 70K. However, the SPE emits until much higher temperatures showing strong localization requiring much higher temperature to thermalize. Interestingly, the emergent emission from LEs are prominent below 40K which happens to coincide with the hidden-order transition[4,5] temperature reported for CrSBr.

# V. Charge transfer interaction from in-plane field measurements

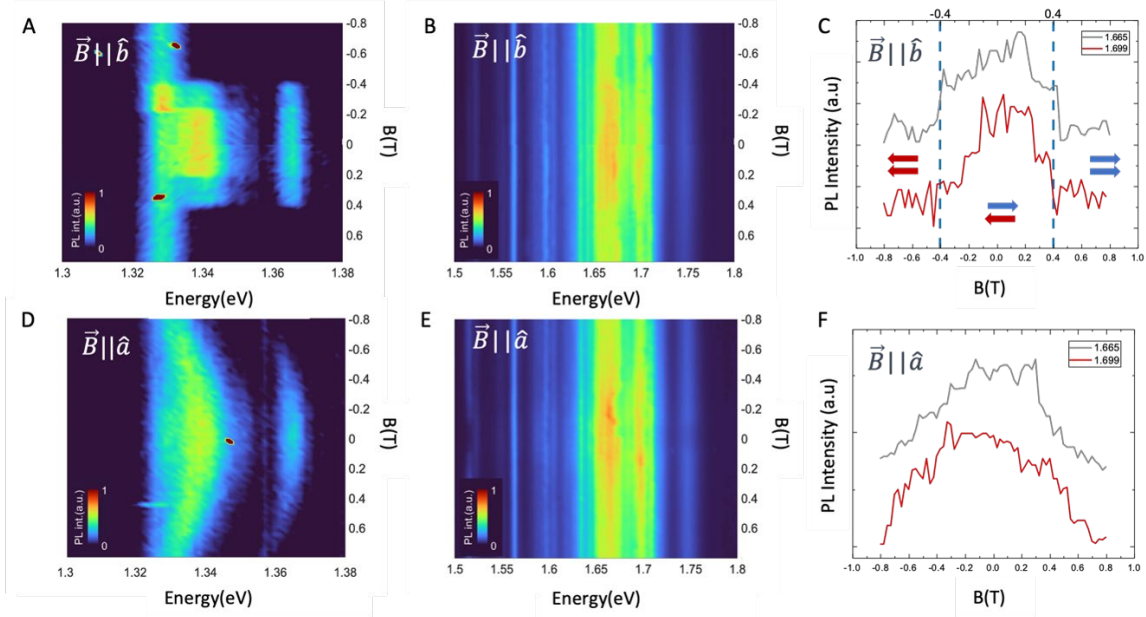

**Figure SI 6**. In-plane magnetic field dependent **μ**-PL measurements for the heterostructure. (a) and (b) Colour plot showing the magnetic field dependence of the CrSBr and WSe$_2$ PL contributions (respectively) to the overall emission from the heterostructure region, with field applied along the easy($\hat{b}$) axis of CrSBr. The colour bar represents the integrated intensity of the PL signal. The two contributions are plotted separately for better comparison as their intensities are very different. (c) The magnetic field dependence of the intensities of the two main bands of WSe$_2$ PL at 1.665eV (grey) and 1.699eV (red) for field applied along the easy($\hat{b}$) axis of CrSBr. (d) and (e) show the same measurements as in (a) and (b) with the field oriented in the intermediate($\hat{a}$) axis of CrSBr. (f) Similar plot as (c) for the field oriented in the intermediate($\hat{a}$) axis of CrSBr. The dotted line marks the transition fields.

This result is very similar to the recently reported MoSe$_2$/CrSBr heterostructure where a charge transfer interaction due to a broken type III band alignment is proposed to be responsible for the coupling of MoSe$_2$ PL intensity to the magnetic state of CrSBr.[6] The electronic structure of the two magnetic states of CrSBr are quite different,[6] hence it is not surprising that the interaction with the contiguous layers is also altered by the phase transition. One can take advantage of this effect to monitor the magnetic state of CrSBr from the WSe$_2$ emissions, which is particularly useful for few-layer flakes of CrSBr (as in our case) as their PL emission is much weaker compared to that of WSe$_2$ monolayer.